\documentclass[preprint,showpacs,preprintnumbers,amsmath,amssymb,floats, aps]{revtex4}

\usepackage{blindtext}
\usepackage{graphicx}
\usepackage[english]{babel}
\usepackage{amsmath}
\usepackage{amssymb}
\usepackage{color}
\def\AA{\mathring{\mathrm{A}}}

\begin{document}

\title{Time-dependent correlations of the Edwards-Anderson order parameter above the spin-glass transition}

\author{
Jingjin Song ${ }^{1}$, Sheena K.K. Patel ${ }^{1,2}$, Rupak Bhattacharya ${ }^{1}$, Yi Yang ${ }^{1}$, Sudip Pandey ${ }^{1}$, Xiao M. Chen ${ }^{3}$, Eric Lee-Wong ${ }^{1}$, Kalyan Sasmal ${ }^{1}$, M. Brian Maple ${ }^{1}$, Eric E. Fullerton ${ }^{2}$, Sujoy Roy ${ }^{3}$, Claudio Mazzoli ${ }^{4}$, Chandra M. Varma${ }^{5}$ and Sunil K. Sinha* ${ }^{1}$}
\vspace{-0.5 cm}
\affiliation{
[1] Department of Physics, University of California San Diego, La Jolla, CA 92093-0319\\
[2] Center for Memory and Recording Research, University of California San Diego, La Jolla, CA 92093-0401\\
[3] Advanced Light Source, Lawrence Berkeley National Laboratory, Berkeley, CA 94720\\
[4] National Synchrotron Light Source II, Brookhaven National Laboratory, Upton, New York, 11973-5000\\
[5] Dept. of Physics, University of California Berkeley, Berkeley, CA 94720}

\begin{abstract}
In  1975 Edwards and Anderson introduced a new paradigm  that interacting quenched systems, such as a spin-glass, have a phase transition in which long time memory of spatial patterns is realized without spatial correlations.  We show here that the information about the time-dependent  correlations above the spin-glass transition are embedded in  the four spin  correlations of the intensity of speckle pattern.  This encodes the spin-orientation memory and  can be measured by the technique of resonant magnetic x-ray photon correlation spectroscopy (RM- XPCS). We have implemented this method to observe and accurately characterize the critical slowing down of the spin orientation fluctuations in the classic metallic spin glass alloy $Cu_{1-x}{Mn}_x$ over time scales of ${2}$ sec. to $2 \times 10^{\mathbf{4}}$ secs. Remarkably the divergence of the correlation time as a function of temperature is consistent with the Vogel-Vulcher law,  universally used to characterize the viscous relaxation time in structural  glasses.  Our method also opens the way for studying phase transitions in systems such as spin ices, quantum spin liquids, the structural glass transition, as well as possibly provide new perspectives on the multifarious problems in which spin-glass concepts have found applications.
\end{abstract}
\maketitle

The specific realization that a spin-glass encodes long-term memory of spin-orientations below a transition temperature $T_g$ came through the introduction of a novel class of  order parameters [1], [2].  The inspiration for this were magnetic susceptibility experiments in metals with magnetic impurities with spins $S_{i}$ located randomly at sites $i$. A cusp was found in the temperature dependence of the magnetic susceptibility of such "spin-glasses" (SG) which become sharper as the frequency of measurements and the magnetic field is reduced towards zero [3-7].

The multiple states possible in a spin-glass with even more multiple routes of connection between such states has found parallels with theories of protein folding, prebiotic evolution, neural networks, stochastic linear programming problems, machine learning and other problems in computational sciences [7-11].

While the SG problem has been mathematically fecund [2] and with many off-shoots,  few experimental methods have given information on the microscopic correlation functions either in the spin-glass phase or in the transition to it. Macroscopic experiments noted above have been of course very instructive. There were early inelastic neutron scattering studies of spin-glasses which studied the two spin correlations as the SG transition was approached, similar to the study of critical spin-fluctuations near the transition to ordered states in ferromagnets or anti-ferromagnets, but these did not show the dynamics of the usual critical phenomena [12-14]. Moreover, the time scales and the nature of fluctuations in SGs are radically different from the typical time scales probed with neutron scattering (including the neutron spin-echo (NSE) technique [15,16]), as we show below. 

 Since the order parameter of spin-glasses is itself a correlation of a pair of spins [1], only specific four spin correlations can give information on
the evolution in time of the fluctuations of the actual order parameter above the SG ordering temperature $T_g$ .
 To facilitate the discussion, let us introduce 
\begin{equation*}
q(t, \tau)=\frac{1}{N} \sum_{i}\int \frac{d\Omega_i}{4 \pi} \left\langle {\bf S}_{i}(t)\right\rangle_{T}\left\langle {\bf S}_{i}(t+\tau)\right\rangle_{T}. \tag{1}
\end{equation*}
$\Omega_i$ is the two-dimensional angle that the classical vector ${\bf S}_i$ makes with respect to a fixed axis.  $q(t, \tau)$ gives the time $\tau$ over which memory of the pattern of randomness in the frozen configuration is retained through  the overlap of the thermal average of the orientation of $\left\langle {\bf S}_{i}(t)\right\rangle_{T}$ of a spin located at a site $i$ with the same object at a later time $(t+\tau)$. Only then is the average over the spins at their random locations taken.
In Eq. (1),  $N$ is the total number of spins, the subscript $T$ is the thermal average (henceforth the subscript $T$ will be dropped) and the sum over the random location of spins $i$ effectively averages over randomness {\it after} the temporal correlations are evaluated. The Edwards-Anderson (EA) order parameter [1] is given by
\begin{equation*}
q_{EA} = Lim_{\tau \rightarrow \infty}\langle q(t, \tau)\rangle_{t} \tag{1}
\end{equation*}
As expected, we show below that the experiments reveal no observable dependence of $q(t, \tau)$ on $t$. $q(\tau)$, as shown below, is  related to the generalization to finite $\tau$ of the spin-glass susceptibility $\chi_{SG}$ defined earlier [6]. 

Here, first we show that the temporal correlations in the intensity of Resonant Magnetic X-ray Photon Correlation Spectroscopy (RM-XPCS) can be used to study the dynamical critical behavior of a spin-glass by directly probing,   as their principal contribution, $q(t, \tau)$. These are measured in real time on time scales much longer than those available to NSE or any other inelastic scattering technique. The results show very slow decay of the fluctuation relaxation time with temperature, so that these fluctuations are still evident above 1.5 $T_g$, far above the critical regime of ordinary continuous transitions. We then describe the results of the measurements and their interesting conclusions.

{\it Correlations in XPCS Measurements}: 

Resonant magnetic scattering (usually with the X-ray photons tuned to the L- or M-edge of the magnetic atoms) gives a large enhancement of the scattering amplitude due to the magnetic moment at resonance. If the photon energy is off-resonance, the scattering is only from charge scattering. The recorded signal on the detector consisted of speckle patterns [19-22] consisting of intensity fluctuations arising from the random phase interferences of the scattering from all the ions from the coherent X-ray beam. These consist of both small-angle charge scattering (which is static on the time scales of these measurements), and resonant magnetic scattering from the disordered and fluctuating Mn spins.

In the XPCS technique, the time-averaged intensity-intensity autocorrelation function is measured and normalized by the averaged intensity squared, resulting in a second-order correlation function $\mathrm{g}_{2}(\mathbf{q}, \tau)$. This is given by [19-22]
\begin{equation*}
\mathrm{g}_{2}(\mathbf{q}, \tau)=\frac{\left\langle I_{t}(\mathbf{q}, t) I_{t}(\mathbf{q}, t+\tau)\right\rangle_{\mathrm{t}}}{\left\langle I_{t}(\mathbf{q}, t)\right\rangle_{\mathrm{t}}^{2}}. \tag{3}
\end{equation*}
\noindent
 $\langle\ldots\rangle_{t}$ represents an average over $t$, and $\mathrm{I}_{\mathrm{t}}(\mathbf{q}, \mathrm{t})$ represents the total (charge (c) + magnetic (m) ) scattered intensity at the wave vector transfer $\mathbf{q}$:
 \begin{equation*}
I_{t}(\mathbf{q}, t)=I_{c}(\mathbf{q}, t)+I_{m}(\mathbf{q}, t).\tag{4}
\end{equation*}
The magnetic signal is not large even in the vicinity of the resonance X-ray energy, because of the low concentration of Mn ions, and the charge scattering dominates at small values of the wave vector transfer $\mathbf{q}$ and decreases rapidly with its magnitude. By comparing the resonant and off-resonant scattering we determined that in our sample the magnetic scattering dominates for $|{\bf q}| > 0.005 \AA^{-1}$. Since the charge scattering and the magnetic dipole scattering are polarized perpendicular to each other, they do not interfere with one another, so that it is not possible to heterodyne off the (static) charge scattering. Instead the latter acts as a static background under the signal, so the RM-XPCS analysis has to be slightly modified, as discussed below. In the dipolar approximation for quasi-elastic resonant magnetic scattering [17,18] $\mathrm{I}_{\mathrm{m}}(\mathbf{q}, \mathbf{t})$ is given by
\begin{equation*}
I_{m}(\mathbf{q}, t)=\frac{1}{N} C \sum_{i j}\left\langle\left(\hat{\boldsymbol{e}}_{i n} \times \hat{\boldsymbol{e}}_{o u t}\right) \cdot \boldsymbol{S}_{i}(t)\left(\hat{\boldsymbol{e}}_{i n} \times \hat{\boldsymbol{e}}_{o u t}\right) \cdot \boldsymbol{S}_{j}(t)\right\rangle e^{-i \mathbf{q} \cdot\left(\boldsymbol{R}_{\boldsymbol{i}}-\boldsymbol{R}_{\boldsymbol{j}}\right)} \tag{5}
\end{equation*}
if we keep only the dominant linear term in the magnetic spins in the X-ray magnetic cross-section. In Eq. (5), $\langle\ldots\rangle$ stands for thermal average at time t ; the sum over sites carries out the configuration average; $\hat{\boldsymbol{e}}_{\text {in }}$ and $\hat{\boldsymbol{e}}_{\text {out }}$ represent unit polarization vectors for the incident and scattered photons on the sample respectively, $\mathbf{S}_{\mathrm{i}}$ and $\mathbf{S}_{\mathrm{j}}$ are the spins on lattice sites $\mathbf{R}_{\mathrm{i}}$ and $\mathbf{R}_{\mathrm{j}}$ respectively, and the spin operators are equal time operators whose thermal averages are evaluated at time ${t}$ in a small interval around it, which we take to be $2$ secs. which allows enough intensity to be gathered. Finally, N is the total number of spins in the measuring volume and C is a combination of instrumental and geometrical factors and resonant dipole matrix elements. A speckle is produced when macroscopic number of ${\bf S}_i$ interfere coherently during this time interval. We measure the time-scale up to  which the speckle pattern retains its correlation by measuring the correlation of $I_m({\bf q},t)$ to the same at a later time up to about $2\times10^4$ secs. The spin-glass transition is characterized by the divergence of the correlation time.

 It can easily be shown, that if the scattered beam makes a small angle $\theta$ to the incident beam direction (small angle approximation),
$\left(\hat{\boldsymbol{e}}_{\text {in }} \times \hat{\boldsymbol{e}}_{\text {out }}\right) \cdot \boldsymbol{S}_{i} \cong S_{i}^{Z}$, 
where $S_{\mathrm{i}}^{\mathrm{z}}$ is the component of $\boldsymbol{S}_{\mathrm{i}}$ along the incident beam direction. We neglect corrections of the order $\theta$, which is equivalent to forward scattering ($\mathbf{q} \to 0$ in Eq. (5)), dominantly observed and used in the experiments. Thus, we can write Eq. (5) as
\begin{equation*}
I_{m}(\mathbf{q}, t)=\frac{1}{N} C \sum_{i j}\left\langle S_{i}^{Z}(t)\right\rangle\left\langle S_{j}^{Z}(t)\right\rangle \tag{6}
\end{equation*}
\noindent
where the brackets represent thermal averages of the spin operators at the time $t$. In our experiment, we found, as shown in the Supplementary (S)-section F,  that the above correlation function, averaged over long times $t$, indeed showed very little q -dependence at small q.  Then the autocorrelation function for the magnetically scattered intensities can be approximated by
\begin{align*}
& \left\langle I_{m}(\mathbf{q}, t) I_{m}(\mathbf{q}, t+\tau)\right\rangle_{t} 
 \approx \frac{1}{N^{2}} C^{2} \sum_{i j}\left\langle\left\langle S_{i}^{Z}(t)\right\rangle\left\langle S_{j}^{Z}(t)\right\rangle\left\langle S_{i}^{Z}(t+\tau)\right\rangle\left\langle S_{j}^{Z}(t+\tau)\right\rangle\right\rangle_{t} \\
& \quad+\frac{1}{N^{2}} C^{2} \sum_{i j}\left\langle\left\langle S_{j}^{Z}(t)\right\rangle\left\langle S_{j}^{Z}(t)\right\rangle\right\rangle \sum_{k l \neq i j}\left\langle\left\langle S_{k}^{Z}(t+\tau)\right\rangle\left\langle S_{l}^{Z}(t+\tau)\right\rangle\right\rangle_{t} \tag{8}
\end{align*}
\noindent
In the first term, we have kept the autocorrelation over $t$ for the same pairs of spins at times $t$ and ${t}+\tau$, while in the second term we have decoupled the time averages over different pairs of spins because they are assumed spatially
 uncorrelated. By Eq. (7), the second term is equal to $\left\langle\left[I_{m}(\mathbf{q}, \mathrm{t})\right]\right\rangle_{\mathrm{t}}^{2}$ to order ($1/\mathrm{N}$). Thus, we may write Eq. (8) as
\begin{equation*}
\left\langle I_{m}(\mathbf{q}, t) I_{m}(\mathbf{q}, t+\tau)\right\rangle_{t}= C^2\chi_{S G}(\tau)+\left\langle I_{m}(\mathbf{q}, t)\right\rangle_{t}^{2} \tag{9}
\end{equation*}
\noindent
where $\chi_{SG}(\tau)$  in the first term in Eq. (9)  is the generalization to finite $\tau$ of the usually defined spin-glass susceptibility [6]. Comparing with Eq. (1), it is related to the fluctuations of the EA order parameter,
\begin{align*}
\chi_{S G}(\tau)
=\frac{1}{N^{2}} \left\langle q(t, \tau)^{2}\right\rangle_{t}. \tag{10}
\end{align*}
so that below the transition at $\tau \to \infty$, it is proportional to $\mathrm{q}_{EA}^{2}$. The factor C contains the solid angle $\Delta \Omega$ subtended by the group of pixels contributing to $\mathrm{I}_{\mathrm{m}}(\mathbf{q}, \mathrm{t})$ so that the quantity above is of order $(\Delta \Omega / 4 \pi)^{2}$.

In XPCS, the total intensity autocorrelation function $\mathrm{g}_{2}(\mathbf{q}, \tau)$, can be written as
\begin{equation*}
\mathrm{g}_{2}(\mathbf{q}, \tau)=1+\beta \frac{\left\langle I_{t}(\mathbf{q}, t) I_{t}(\mathbf{q}, t+\tau)\right\rangle_{t}-\left\langle I_{t}(\mathbf{q}, t)\right\rangle_{t}^{2}}{\left\langle I_{t}(\mathbf{q}, t)\right\rangle^{2}} \tag{11}
\end{equation*}
\noindent
where $\beta$ is the contrast factor [19-22] arising from the partial coherence of the X-ray beam. Using Eq. (4), and the fact that the charge scattering is independent of time, and that there is no interference between charge and magnetic scattering, we obtain
\begin{equation*}
\mathrm{g}_{2}(\mathbf{q}, \tau)=1+\beta \frac{\left\langle I_{m}(\mathbf{q}, t) I_{m}(\mathbf{q}, t+\tau)\right\rangle_{t}-\left\langle I_{m}(\mathbf{q}, t)\right\rangle_{t}^{2}}{\left\langle I_{t}(\mathbf{q}, t)\right\rangle_{t}^{2}} =1+\beta \frac{\chi_{S G}(\tau)}{\left\langle I_{t}(\mathbf{q}, t)\right\rangle_{t}^{2}} \tag{12}
\end{equation*}
where in the second equality, we have used Eq. (9).
Thus, the function $\left[\mathrm{g}_{2}(\mathbf{q}, \tau)-1\right]$ measures the 4 -spin correlations as given in Eq. (10) above. In presenting the experimental results we remove the inessential non-universal experimental features, and plot the normalized quantity
\begin{equation*}
\mathrm{\tilde{g}}_2(\mathbf{q}, \tau)  \equiv (\mathrm{g}_{2}(\mathbf{q}, \tau) - 1)/(\mathrm{g}_{2}(\mathbf{q}, 0) - 1). \tag{13}
\end{equation*}

{\it Experimental Procedures}:
\noindent
The measurements were carried out on a $\mathrm{Cu}_{0.88} \mathrm{Mn}_{0.12}$ alloy film. The details of the sample and the experimental details of the RM-XPCS are given in
Supplemental information (Supp.) ${\bf A}$. 

\begin{figure}[h]
 \begin{center}
 \includegraphics[width= 0.8\columnwidth]{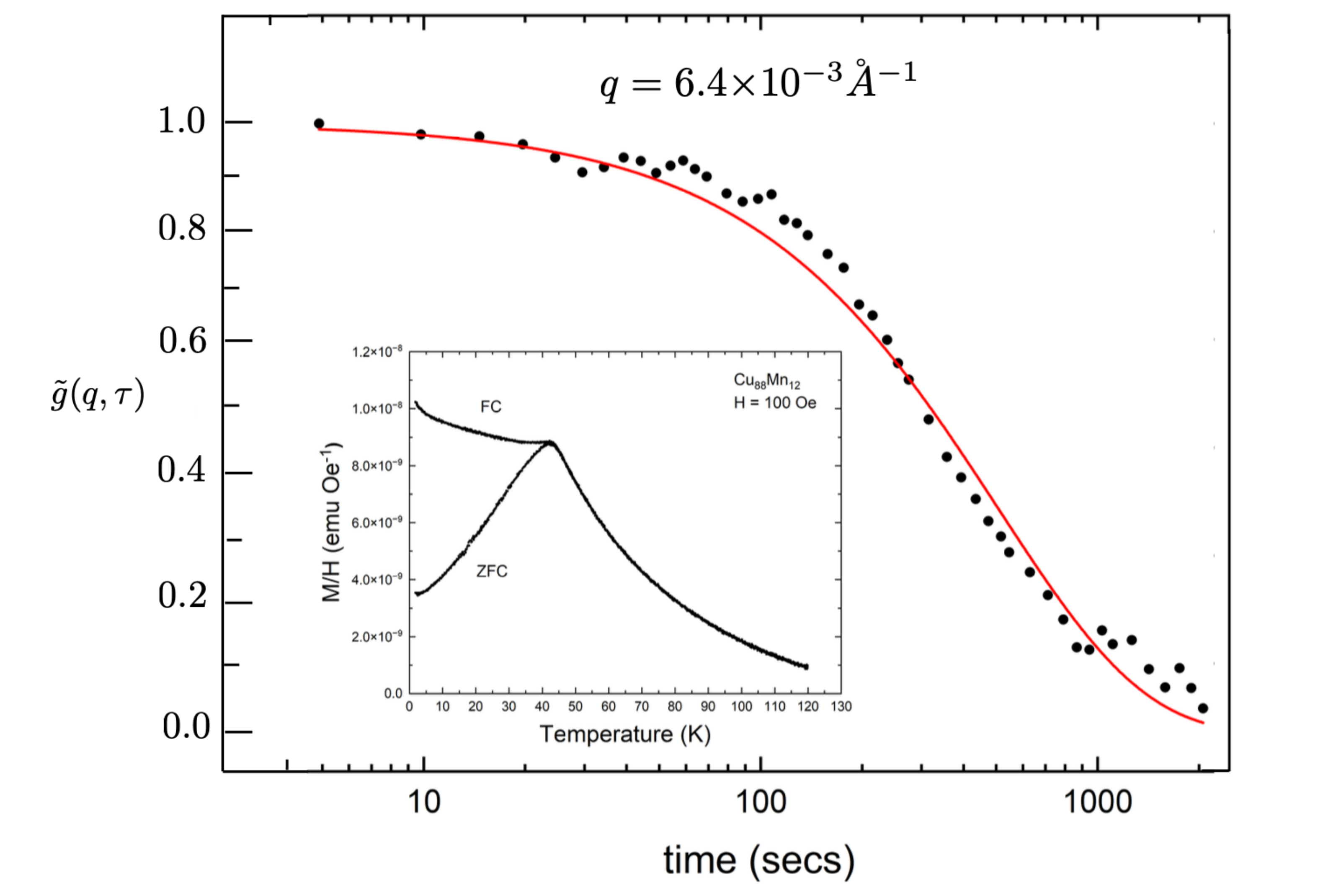}
 \end{center}
\caption{Inset: Measured field-cooled and zero-field-cooled dc susceptibilities of the $\mathrm{Cu}_{0.88} \mathrm{Mn}_{0.12}$ sample as a function of temperature.
The main figure shows experimental data for the normalized function $\left[\mathrm{\tilde{g}}_{2}(\mathbf{q}, \tau)\right]$ at $q=6.4 \times$$10^{-3} \AA^{-1}$ at $\mathrm{T}=74 \mathrm{~K}$ obtained from RXPCS, (curve has been normalized to 1 at $\mathrm{t}=0$.) The red-curve is Eq. (14) from which $\tau_0(T)$ is deduced.
}
 \label{Fig:SG1} 
\end{figure}

Low field (100 Oe) dc magnetic susceptibility measurements as a function of temperature (inset in Fig. (\ref{Fig:SG1}) showed typical SG behavior for both field-cooled and zero-fieldcooled measurements, with an estimated $\mathrm{T}_{\mathrm{g}}$ of 45 $\pm$ 2K .  
We present  our experimental results from XPCS and consistency checks for measurements of $\mathrm{g}_{2}(\mathbf{q}, \tau)$ as follows:\\
(a) We verified that $\mathrm{\tilde{g}}_{2}$ was independent of the starting time $\mathrm{t}$ for measuring the time dependent intensity autocorrelation function -see Supp. {\bf C}.\\
(b) In order to take into account the possible intensity fluctuations of the incident beam, the $\mathrm{\tilde{g}}_{2}$ functions were corrected by dividing by the time autocorrelation function of the incident beam intensity - See Supp. {\bf D}.\\
(c) The off-resonance measurements (pure charge scattering) at all $q$-values, and the on resonance measurements of $\mathrm{\tilde{g}}_{2}$ at small q showed no time dependence, as expected. The larger $\mathrm{q} (\gtrsim 0.005 \AA^{-1})$ where the spin scattering dominated, showed time dependence, but there was very little dependence of the magnetic scattering contribution to $\mathbf{g}_{2}(\mathbf{q}, \tau)$ on $\mathbf{q}$, over the range of wavevectors studied. The near independence with ${\bf q}$ is shown in the supplementary section in Supp. {\bf E}- Fig.(3).\\
(d) The contrast factor $\beta$ as measured by Eq. (11) from the un-normalized curves for $\mathrm{\tilde{g}}_{2}(\mathbf{q}, \tau)$ came out to be of order $10^{-3}$, typically 2 orders of magnitude smaller than in a conventional XPCS experiment. This can be attributed, in part, to the weakness of the magnetic scattering signal, as well as mainly from two causes: (1) the fact that there are many spin excitations which decay on much faster length scales and whose contributions have vanished before our first measuring time frame is completed (as seen in the NSE experiments [15,16], for example) and (2) because we are left with only same-spin correlation functions in the final time averaged $\mathrm{g}_{2}$ function which reduces the intensity by a factor $1 / \mathrm{N}$ compared to that from normal
correlation functions which involve interference scattering from different spins. The situation is somewhat reminiscent of incoherent vs. coherent neutron scattering from assemblies of atoms.

{\it Experimental Results}:

The principal results are the following: The normalized time-dependence in  $\mathrm{\tilde{g}}_{2}$  can be fitted very well  at all temperatures measured by the form
\begin{equation*}
\mathrm{\tilde{g}}_{2}(\boldsymbol{q}, \tau)= \frac{C_1 + C_2e^{-\left(\frac{\tau}{\tau_{0}}\right)}}{C_1 + C_2} \tag{14}
\end{equation*}
\noindent
where the relaxation time $\tau_{0}(\mathrm{~T})$ shows little dependence on q-values but is temperature dependent. This is shown at $T=74 K$ and at the specified $q$ in Fig. (\ref{Fig:SG1}).
The near independence of the $q$ dependence is exhibited in Supp. {\bf E}.

 \begin{figure}[h]
 \begin{center}
 \includegraphics[width= 0.8\columnwidth]{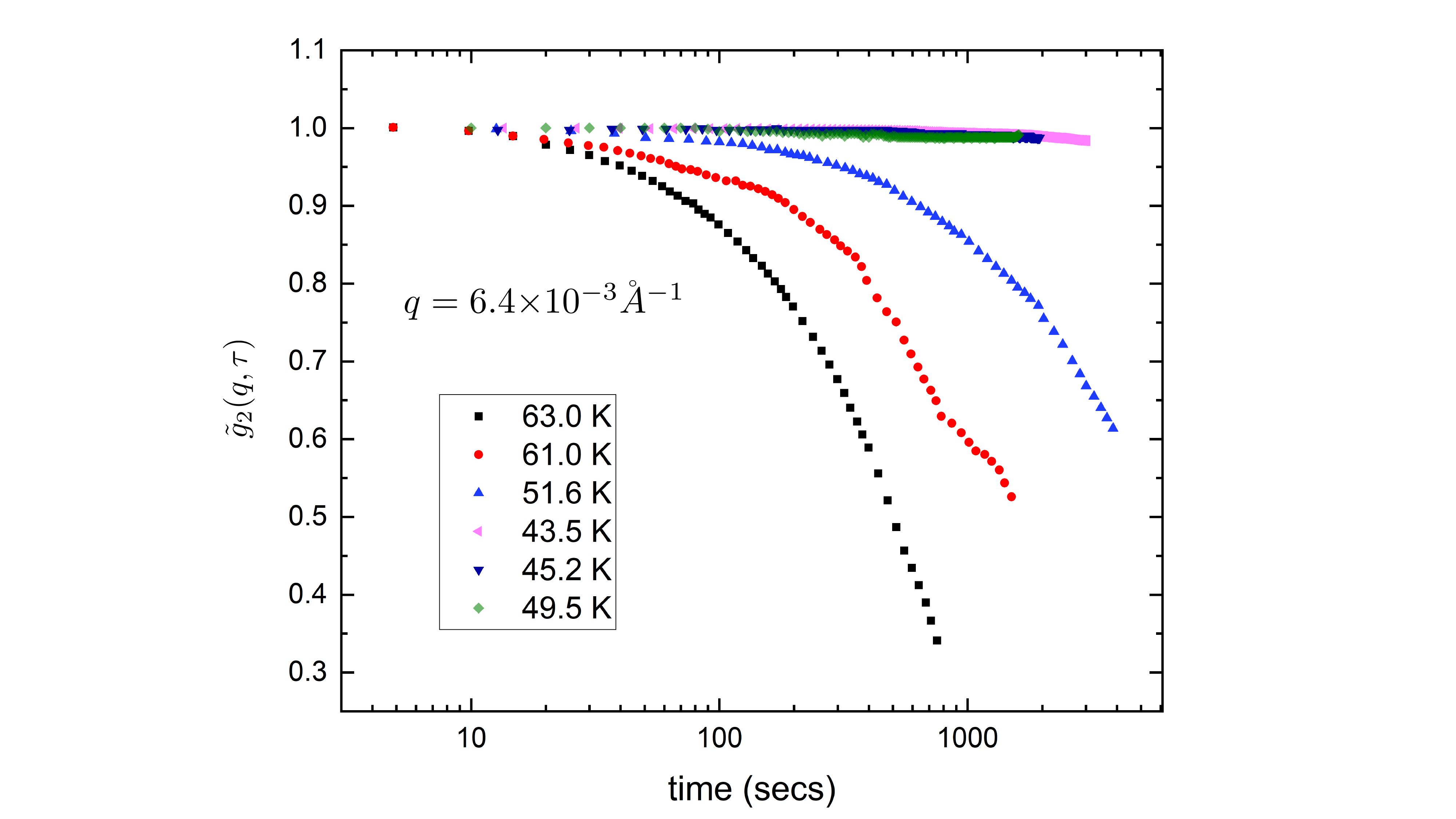}
 \end{center}
 \caption{The function $\mathrm{\tilde{g}}_{2}(q, \tau)$ at $q = 6.4 \times 10^{-3} \AA^{-1}$ for several different temperatures. As already mentioned, these curves have all been normalized to unity at $\mathrm{\tau}=0$.}
 \label{Fig:SG2} 
\end{figure}

Fig. (\ref{Fig:SG2}) shows how the relaxation time increases rapidly as the temperature is cooled towards T$_g$. However, there is a limiting time of $\sim 4,000$ secs, because of the decay in time of the autocorrelation function of the incident beam intensity from the synchrotron itself. This was taken into account  by normalizing the $\mathrm{g}_{2}$ functions by those of the incident beam, but it limits our effective measurements of the spin dynamics to time scales of $\lesssim 2 \times 10^{4}$ secs
with acceptable error bars. 

\begin{figure}[h]
 \begin{center}
 \includegraphics[width= 0.8\columnwidth]{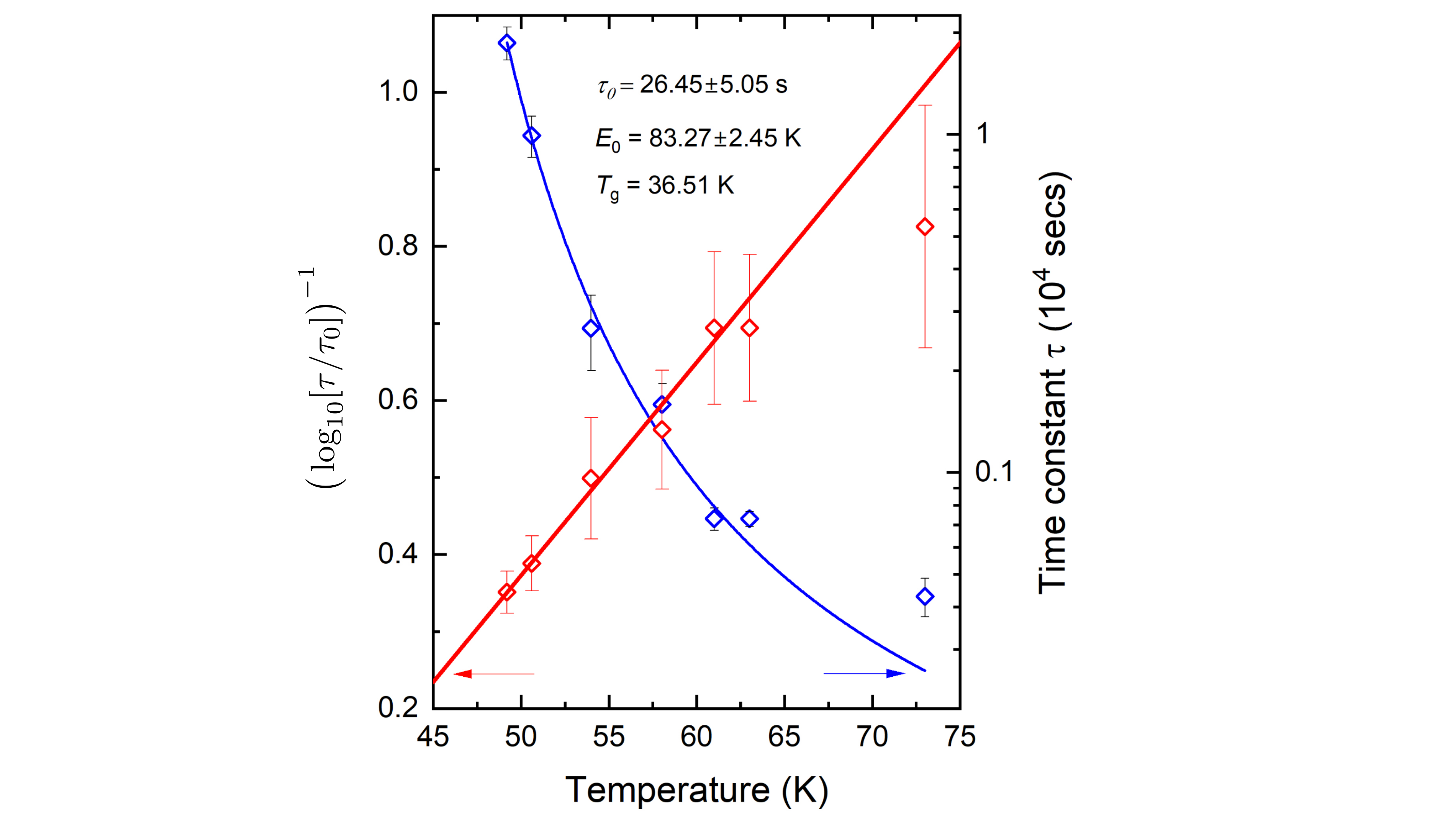}
 \end{center}
\caption{The deduced relaxation time as a function of temperature is plotted in two different ways. The blue data points shows  $\tau(T)$ on a logarithmic scale and the blue curve is the fit to Eq. (15). The same data  is shown in red drawn with the vertical axis on the left so that the Vogel-Fulcher law should appear as linear in $(T-T_0)/E_0$, as in the red line. }
 \label{Fig:SG3} 
\end{figure}

The measured temperature dependence of $\tau_0$ is shown in Fig. (\ref{Fig:SG3}). It is remarkable that the temperature dependence of the relaxation rate fits very well the Vogel-Fulcher form,
\begin{equation*}
\tau_0(T)=\tau_{0} \exp \left(\frac{E_{0}}{T-T_{0}}\right) \tag{16}
\end{equation*}
\noindent
familiar in the discussion of the glass transition [27,28]. The fit to the Vogel-fulcher form is shown in Fig.(\ref{Fig:SG3}) in two ways, the direct dependence of $\tau(T)$, and as $\big(\log_{10} (\tau/\tau_0))^{-1}$ which is proportional to $(T-T_0)/E_0$ if Eq. (16) is a valid description. This form has been derived from the shear correlations of the structural glass [29], showing its connection to  the linear in $T$ specific specific heat observed in all glasses.
The linear in $T$ specific heat is also found in spin-glasses [30]. The fit $\tau(T)$ is very good, especially at the more important longer times. The corresponding value of $\mathrm{T}_{0}$ is 36.5 K which is about 5 K below the value where the susceptibility shows a cusp. It is significant that the $T_0$ in the Vogel-Fulcher law for structural glasses is actually the extrapolated temperature, often called the Kauzmann temperature \cite[31, 32], where the configurational entropy of the liquid goes to $0$ [31]. The Kauzman temperature is unobservable because the
viscous relaxation rate becomes un-observable below a temperature referred to as the glass transition temperature $T_g$ which is always larger than $T_0$. 
The similarity of the measured relaxation time $\tau_0(T)$ in spin-glasses with the viscous relaxation time in structural glasses calls for a systematic explanation. In latter, the Vogel-Fulcher law is also derived from replica based formalism of spin-glasses supplemented by  random fields [33]. For spin-glasses no measurements of the correlation time and no derivation of such a form exists.

The data can also be well fitted with a power law:
\begin{equation*}
\tau_{0}(T)=\frac{A}{\left(T-T_{\mathrm{g}}\right)^{B}} \tag{15}
\end{equation*}
\noindent
This fit is shown in the supplementary section in Fig. (5). The fit yields values of 44  K for $\mathrm{T}_{\mathrm{g}}$ and $\mathrm{B}=2.7$. The value of B shows that the dynamical critical fluctuations we observe decay much more slowly than expected from the dynamical critical exponent (zv) of $\sim 7$ observed in earlier computer simulations [23] and nonlinear susceptibility measurements [24]. On the other hand, the mean field value of (zv) is 2 and values even lower have been quoted experimentally [25,26].


These measurements constitute the first direct measurement of the temperature dependence of the time-dependent susceptibility related to the $\mathrm{SG}(\mathrm{EA})$ order parameter fluctuations, and show slowing down of the fluctuations as $\mathrm{T}_{\mathrm{g}}$ is approached from above, consistent with the Vogel-Fulcher law of structural glasses.


Interestingly, the present technique could also be potentially useful in studying the fluctuations of entangled singlet spin pairs that have been proposed for the case of the Resonating valence bond state in quantum spin liquids [34, 35], or the slow fluctuations in spin ices $[36, 37]$, since these would also involve 4 -spin correlation functions. Although the present method concentrates on very slow fluctuations, much faster fluctuations can also be studied by the RM-XPCS technique at time scales of nano seconds or less by using the delayed 2-pulse speckle contrast measurement methods that have recently been developed for free-electron X-ray laser sources [38, 39,40].

\section*{Acknowledgements}
We wish to acknowledge the help from the staff of the CSX beamline at NSLS II, and staff members at ALS in the experiments at the 12.0.2 coherent beamline at ALS. We thank Sheng Ran for assistance with magnetic measurements made at UCSD. The research was supported by the US Department of Energy, Office of Science, Basic Energy Sciences, under Grant no. DE-SC0003678. MBM acknowledges the support of the US Department of Energy, Office of Science, Basic Energy Sciences, under Grant No. DE-FG02-04ER46105. This research used beamline 23-ID-1of the National Synchrotron Light Source II, a US Department of Energy (DOE) Office of Science User Facility operated for the DOE Office of Science by Brookhaven National Laboratory under Contract No. DE-SC0012704. Work at the ALS, LBNL was supported by the Director, Office of Science, Basic Energy Sciences, of the US Department of Energy (Contract No. DE-AC0205CH11231).



\section*{References}
\noindent
[1] Edwards S. F. \& Anderson, P. W. Theory of spin glasses. Journal of Physics F: Metal Physics 5, 965-974 (1975).\\[0pt]
[2] For a compendium of references to early theoretical directions pursued, see D. Sherrington and S. Krikpatrick, 50 years of spin glass theory, arXiv.org/2505.24432 \\[0pt]
[3] Mydosh, J. A. Spin glasses: redux: an updated experimental/materials survey. Rep. Prog. Phys. 78052501 (2015).\\[0pt]
[4] Mydosh, J. A. Spin glasses - recent experiments and systems. Journal of Magnetism and Magnetic Materials 7 237-248 (1978).\\[0pt]
[5] Huang, C. Y. Some experimental aspects of spin glasses: A review. Journal of Magnetism and Magnetic Materials 51 1-74 (1985).\\[0pt]
[6] Young, A. P. Spin Glasses and Random Fields, (Ed., Series on Directions in Condensed Matter Physics, Vol. 12, World Scientific Singapore, 1998).\\[0pt]
[7] Bryngelson, J. D. \& Wolynes, P. G. Spin glasses and the statistical mechanics of protein folding. Proc. Nat. Acad. Sci. 84, 7524-7528 (1987).\\[0pt]
[8] Mezard, M. Parisi, G. \& Virasoro, M. A. Spin Glass Theory and Beyond (World Scientific Lecture Notes in Physics, Vol. 9, World Scientific Press, Singapore 1986).\\[0pt]
[9] Dotsenko, V. An Introduction to the Theory of Spin Glasses and Neural Networks (World Scientific Lecture Notes in Physics, Vol. 54, World Scientific Press, Singapore 1995).\\[0pt]
[10] Franz, S. Exact solutions for diluted spin glasses and optimization problems. Phys. Rev. Lett. 87, 127209 (2001).\\[0pt]
[11] Monasson, R. Optimization problems and replica symmetry breaking in finite connectivity spin glasses. J. Phys. A: Math. Gen., 31,513-529 (1998).\\[0pt]
[12] Murani, A. P. \& Heidemann, A. Neutron-scattering measurement of the Edwards-Anderson order parameter for a Cu-Mn spin-glass alloy. Phys. Rev. Lett. 41, 1402 (1978).\\[0pt]
[13] Mezei, F. The dynamical behaviour associated with the spin glass transition. J.Mag. Mag. Mat. 31-34, 1327 (1983).\\[0pt]
[14] Mezei, F. \& Murani, A. P. J. Mag. Mag. Mat. 14, 211 (1979).\\[0pt]
[15] Pappas, C., Mezei, F., Ehlers, G., Manuel, P. \& Campbell, I. A., Dynamic scaling in spin glasses. Phys. Rev. B 68, 054431 (2003).\\[0pt]
[16] Pickup R. M., Cywinski, R. Pappas, C., Farago, B. \& Fouquet, P. Generalized spin-glass relaxation. Phys. Rev. Lett. 102, 097202 (2009).\\[0pt]
[17] Hannon, J. P., Trammell, G. T., Blume, M. \& Gibbs, D. X-ray resonance exchange scattering. Phys. Rev. Lett. 61, 1245 (1989).\\[0pt]
[18] Blume, M. Magnetic scattering of x rays. J. Appl. Phys. 57, 3615 (1985).\\[0pt]
[19] Sandy, A. R., Zhang, Q. \& Lurio, L. B. Ann. Rev. Mater. Res. 48,167 (2018).\\[0pt]
[20] Grubel, G. Madsen, A. \& Robert, A. X-ray Photon Correlation Spectrocopy in Soft Matter Characterization ( Springer-Verlag, Chapter 18,2008).\\[0pt]
[21] Sutton, M. A review of X-ray intensity fluctuation spectroscopy. Physique 9, 657-667 (2008).\\[0pt]
[22] Sinha, S. K. Jiang, Z. \& Lurio, L. B. X-ray Photon Correlation Spectroscopy Studies of Surfaces and Thin Films. Adv. Mater. 26, 7764-7785 (2014).\\[0pt]
[23] Ogielski, A. T. Dynamics of three-dimensional Ising spin glasses in thermal equilibrium. Phys. Rev. B 32, 7384 (1985).\\[0pt]
[24] Levy, L. P. Critical dynamics of metallic spin glasses. Phys. Rev. B 38, 4963 (1988).\\[0pt]
[25] Parisi G., Ranieri P., Ricci-Tersenghi F. \& Ruiz-Lorenzo J. J. Mean field dynamical exponents in finite-dimensional Ising spin glass. J. Phys. A, Math. and Gen., 30, 7115-7131 (1997).\\[0pt]
[26] Singh, M. K., Prellier, W., Singh, M. P., Katiyar, R. S. \& Scott, J. F. Spin-glass transition in single-crystal $\mathrm{BiFeO}_{3}$. Phys. Rev B 77, 144403 (2008).\\[0pt]
[27] Langer, J. The mysterious glass transition. Physics Today 60, 2, 8 (2007).\\[0pt]
[28] Dyre, J. C. Colloquium: The glass transition and elastic models of glass-forming liquids. Rev. Mod. Phys. 78, 953-972 (2006).\\[0pt]
[29] Varma, C.M.  Theory of melting of glasses, Philosophical Magazine, 103:17, 1650-1663, (2023).\\[0pt]
[30] Zimmerman, J.E., Hoare, F.E.  Low-temperature specific heat of dilute Cu-Mn alloys. Journal of Physics and Chemistry of Solids, 17. 52-56 (1960) doi:10.1016/0022-3697(60)90174-8 \\[0pt]
[31] Kauzmann, W. Chemical Reviews, 43, 219 (1948). \\[0pt]
[32] J. H. Gibbs and E. A. DiMarzio, Journal of Chemical Physics 28, 373 (1958). \\[0pt]
[33] T. R. Kirkpatrick, D. Thirumalai, and P. G. Wolynes, Phys. Rev. A 40, 1045 (1989).\\[0pt]
[34] Anderson, P. W. Resonating valence bonds: A new kind of insulator? Mat. Res. Bull. 8, 153-160 (1973).\\[0pt]
[35] Shimizu, Y., Miyagawa, K., Kanoda, K., Maseato, M. \& Saoto G. Spin liquid state in an organic Mott insulator with a triangular lattice. Phys. Rev. Lett. 91, 107001(2003).\\[0pt]
[36] Yaraskavitch, L. R. et al., Spin dynamics in the frozen state of the dipolar spin ice material $\mathrm{Dy}_{2} \mathrm{Ti}_{2} \mathrm{O}_{7}$. Phys. Rev. B 85, 020410 (R) (2012).\\[0pt]
[37] Matsuhira, K., Hinatsu, Y. \& Sakakibara, T. Nobel dynamic magnetic properties in the spin ice compound $\mathrm{Dy}_{2} \mathrm{Ti}_{2} \mathrm{O}_{7}$. J. Phys. Condens. Matter 13 L737-L746 (2001).\\[0pt]
[38] Gutt, C. et al., Measuring temporal speckle correlations at ultrafast x-ray sources. Opt. Express 17, 55-61 (2009).\\[0pt]
[39] Seaberg, M. H. et al., Nanosecond x-ray photon correlation spectroscopy on magnetic skyrmions. Phys. Rev. Lett. 119, 067403 (2017).\\[0pt]
[40] Sun, Y. et al., Compact hard x-ray split-delay system based on variable-gap channel-cut crystals. Optics Letters 44, 2582-2585 (2019). \\

\newpage
{\bf Supplementary Sections to \\ "Time-dependent correlations of the Edwards-Anderson order parameter above the spin-glass transition"} \\


\noindent
{\bf A: Sample and Instrumentation}
The sample was a polycrystalline film of the alloy $\mathrm{Cu}_{0.88} \mathrm{Mn}_{0.12}$ of thickness $\sim 400 \mathrm{~nm}$ prepared by co-sputtering Cu and Mn in the proper ratios. This was deposited on a $7.5 \mathrm{~mm} \times 7.5$ $\mathrm{mm}~ \mathrm{SiN} / \mathrm{Si}$ substrate film with a $1 \mathrm{~mm} \times 1 \mathrm{~mm}$ window of SiN of thickness 100 nm, so that a soft X-ray beam could be transmitted through it and through the sample in a transmission geometry scattering experiment. The sample was mounted in a He flow cryostat, initially on beamline 23-ID-1 (CSX) at the NSLS-II Light Source, and subsequently on beamline 12.0.2 at the Advanced Light Source (ALS). These beamlines have the capability of producing a coherent beam of photons by transmission through a $10-\mu \mathrm{m}$-diameter ( $5 \mu \mathrm{~m}$ at ALS) pinhole at photon energies tunable around the $\mathrm{Mn} \mathrm{L}_{3}$ edge at $\sim 641 \mathrm{eV}$. Measurements were made at 636 eV, slightly below the resonant edge energy (to optimize the resonant magnetic dipole scattering, while mitigating the peak absorption at the resonance) and also at 10 eV below this energy to study the non-resonant or purely charge scattering. The incident photon beam was linearly polarized in the horizontal plane. Measurements were made in transmission in the forward direction in small angle geometry. The scattered photons were recorded on a 2D detector.

\newpage
{\bf B: Speckle Pattern}
\begin{figure}[ht]
 \begin{center}
 \includegraphics[width= 0.7\columnwidth]{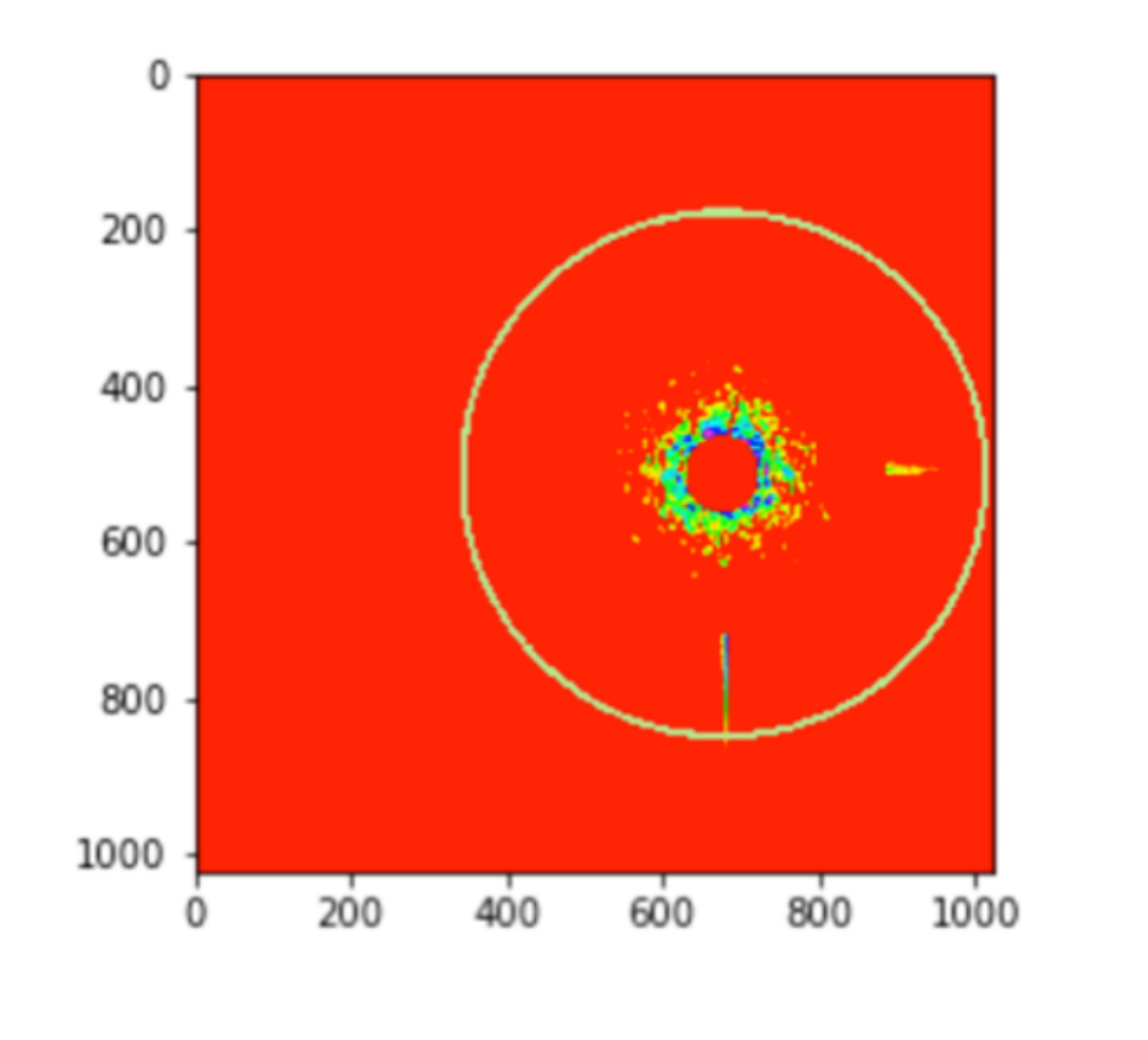}
 \end{center}
 \caption{Typical speckle pattern obtained. The numbers on the axes show pixel number. The circle drawn is at the momentum transfer $\mathrm{q}=6.4 \times 10^{-3} \AA^{-1}$,  to show the scale of momentum transfer vs. the pixel number.}
 \label{Fig:S1} 
 \end{figure}
Typical speckle pattern on the area detector at the ALS in a 5 second exposure at room temperature is shown in Fig. (\ref{Fig:S1}). Speckles due to intense static small angle charge scattering are seen around the beam stop positioned at $\mathbf{q} \approx 0$. Magnetic speckles (typically measured at the larger q -values indicated by the ring at $\mathrm{q}=6.4 \times 10^{-3} \AA^{-1}$, the so-called "high q " values) cannot be seen on this intensity scale, being too weak and rapidly fluctuating.  Streaks seen in image are static artifacts from the sample.

\begin{figure}[h]
 \begin{center}
 \includegraphics[width= 1.0\columnwidth]{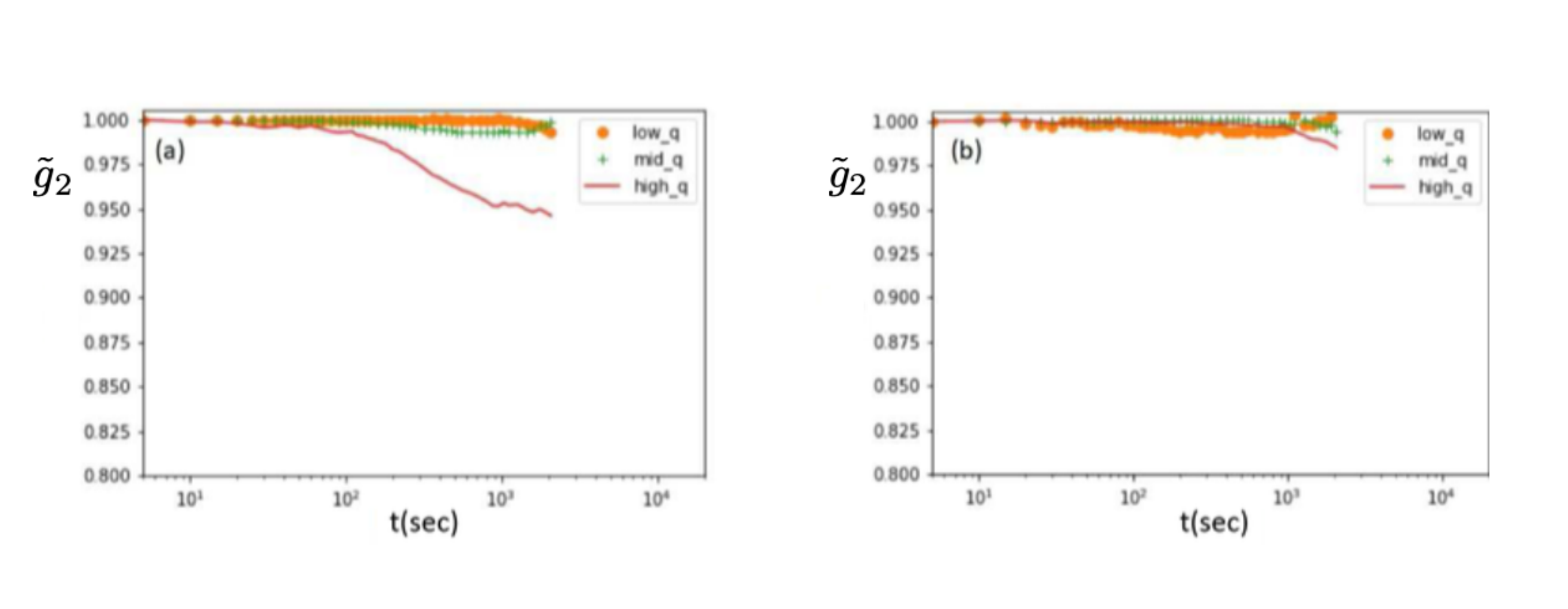}
 \end{center}
 \caption{The function $\mathrm{\tilde{g}}_{2}(q)$ plotted vs. time difference between start and end times for starting times at $\mathrm{t}=0,50$ seconds and 500 seconds respectively, for $\mathrm{q}=6.4 \times 10^{-3} \AA^{-1}$ and 2 different temperatures with the photon beam at the energy of the $\mathrm{Mn} \mathrm{L}_{3}$ resonance energy. The time frame for collecting the time-dependent data was 5 seconds.}
 \label{Fig:S2}
 \end{figure}

 \newpage
 \noindent
{\bf C: Test of independence of normalized $\mathbf{\tilde{g}}_{2}$ function on start times.}\\
By examining $\tilde{g}_{2}(q, t_{2}-t_{1})$ for different starting times $t_{1}$ in specific runs, we verified that it was independent of the starting time $\mathrm{t}_{1}$ and depended only on the time difference, as expected. See Fig. (\ref{Fig:S2}).
 
\begin{figure}[hb]
 \begin{center}
 \includegraphics[width= 0.9\columnwidth]{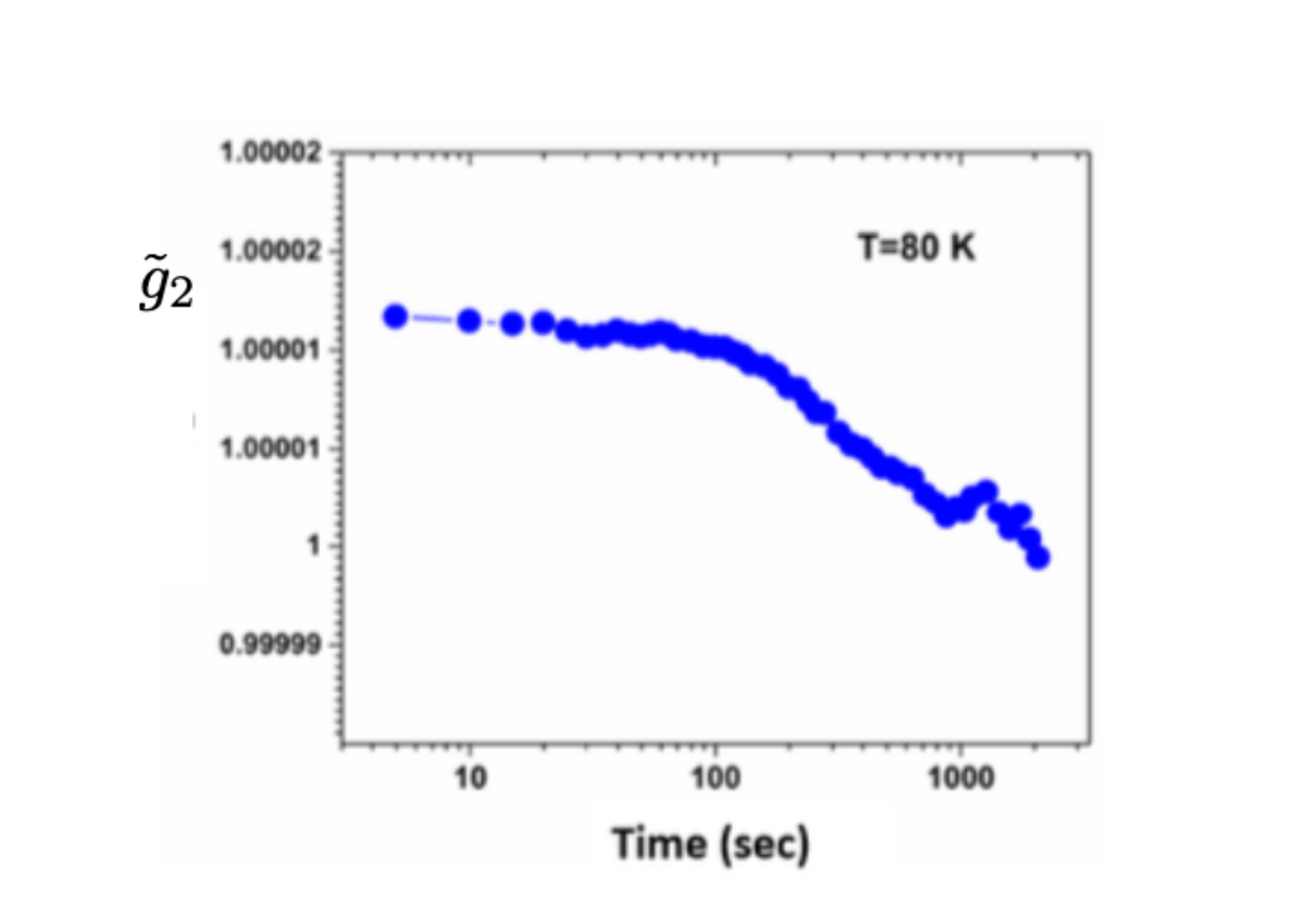}
 \end{center}
\caption{The final $\mathrm{\tilde{g}}_{2}$-functions were divided by the $\mathrm{\tilde{g}}_{2}$-function (self-auto-correlation function) of the incident beam itself to correct for its decay with time. This procedure is based on the fact that the intensity fluctuations of the incident beam and the scattering cross-section are statistically independent. Figure illustrates the $\mathrm{\tilde{g}}_2$-function of the intensity of the main beam, taken as the time-dependent integrated intensity in the 2D detector, as a function of time difference}
 \label{Fig:S3} 
\end{figure}

 \newpage
\noindent
 {\bf D: Corrections for decay of incident beam.}\\
 This is illustrated through Fig.(\ref{Fig:S3}).

\begin{figure}[ht]
 \begin{center}
 \includegraphics[width= 0.9\columnwidth]{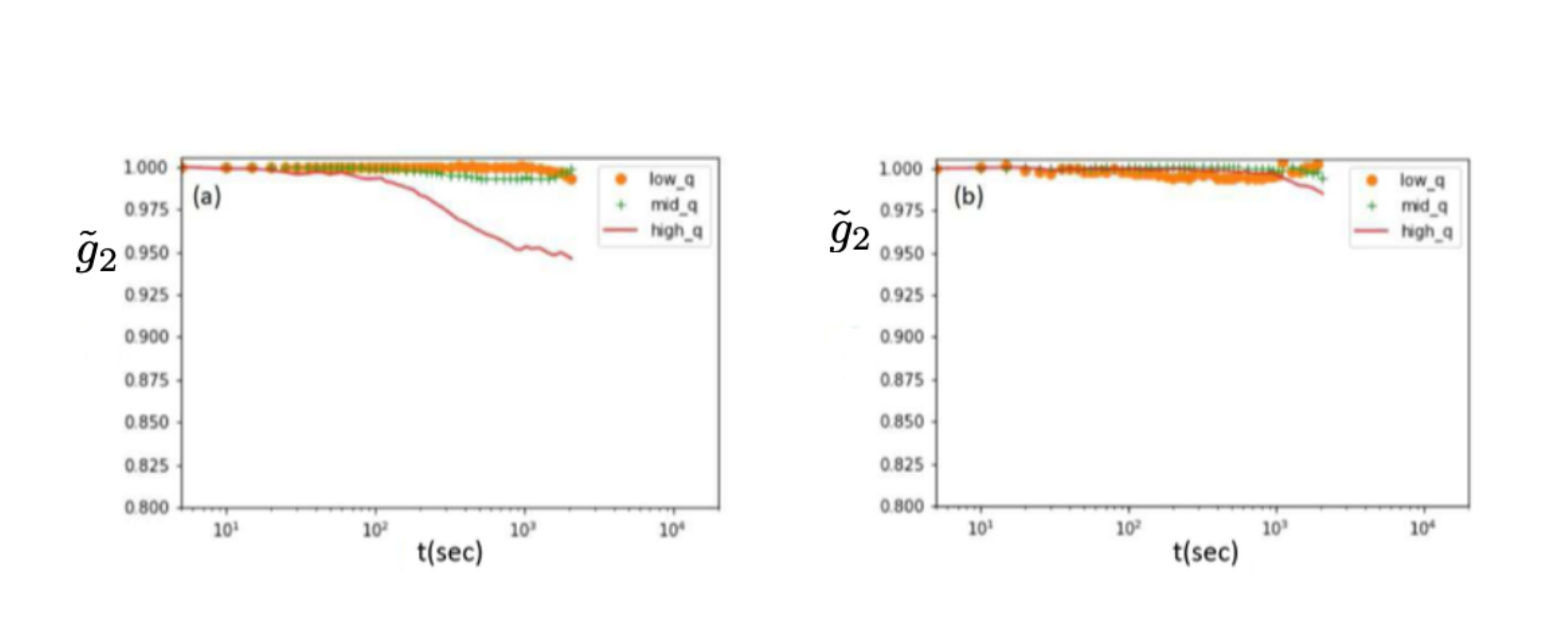}
 \end{center}
\caption{ Plots of $\mathrm{\tilde{g}}_{2}(q)$ vs. time at $T = 80 K$ (a) on resonance at the $\mathrm{Mn} \mathrm{L}_{3}$ edge and (b) off-resonance. The q - values in both cases were: high $\mathrm{q}=6.4 \times 10^{-3} \AA^{-1}$, mid $\mathrm{q}=4.0 \times 10^{-3} \AA^{-1}$, and low $\mathrm{q}=1.9 \times 10^{-3} \AA^{-1}$. At low q -values and at off-resonance, predominantly charge scattering is observed, with no decay of ( $\mathrm{\tilde{g}}_{2}$ ), whereas for larger q and on resonance, predominantly magnetic scattering is observed with concomitant decay of $\left({\tilde{g}}_{2}\right)$.}
 \label{Fig:S4} 
 \end{figure}

\newpage
 \noindent
 {\bf E: Examining the difference between resonant magnetic scattering and charge scattering}\\
 This is shown in Fig. (\ref{Fig:S4}).\\

\begin{figure}[hb]
 \begin{center}
 \includegraphics[width= 0.9\columnwidth]{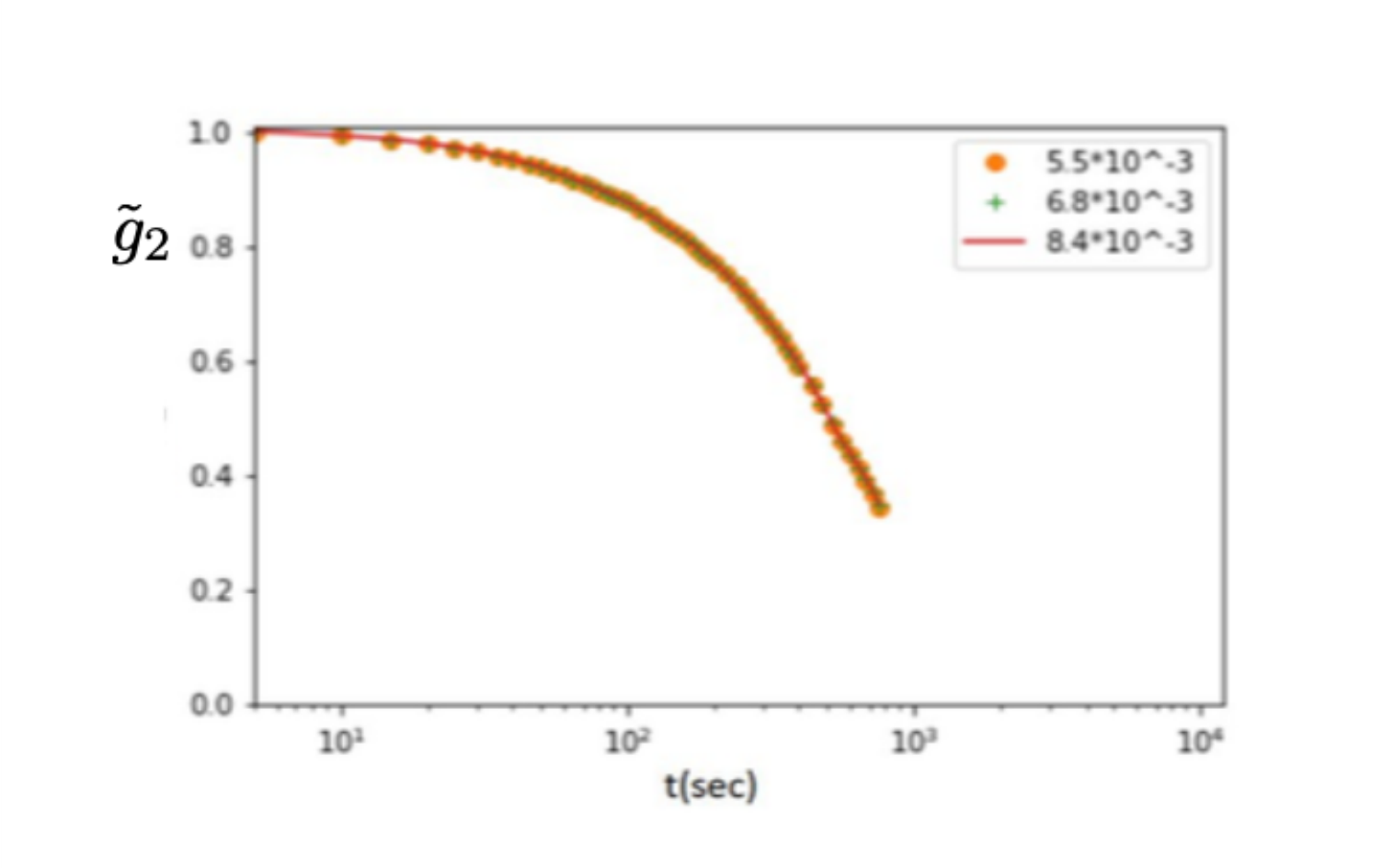}
 \end{center}
\caption{$\mathrm{\tilde{g}}_{2}(q)$ plotted for several q-values in the region where magnetic scattering is dominant. The curves superimpose, showing no q -dependence at these q -values.}
 \label{Fig:S5} 
\end{figure}

\newpage 
\noindent
{\bf F: $q$ independence of $\mathrm{\tilde{g}}_2(q)$}\\
This is shown in Fig. (\ref{Fig:S5}).

\begin{figure}[h]
 \begin{center}
 \includegraphics[width= 0.7\columnwidth]{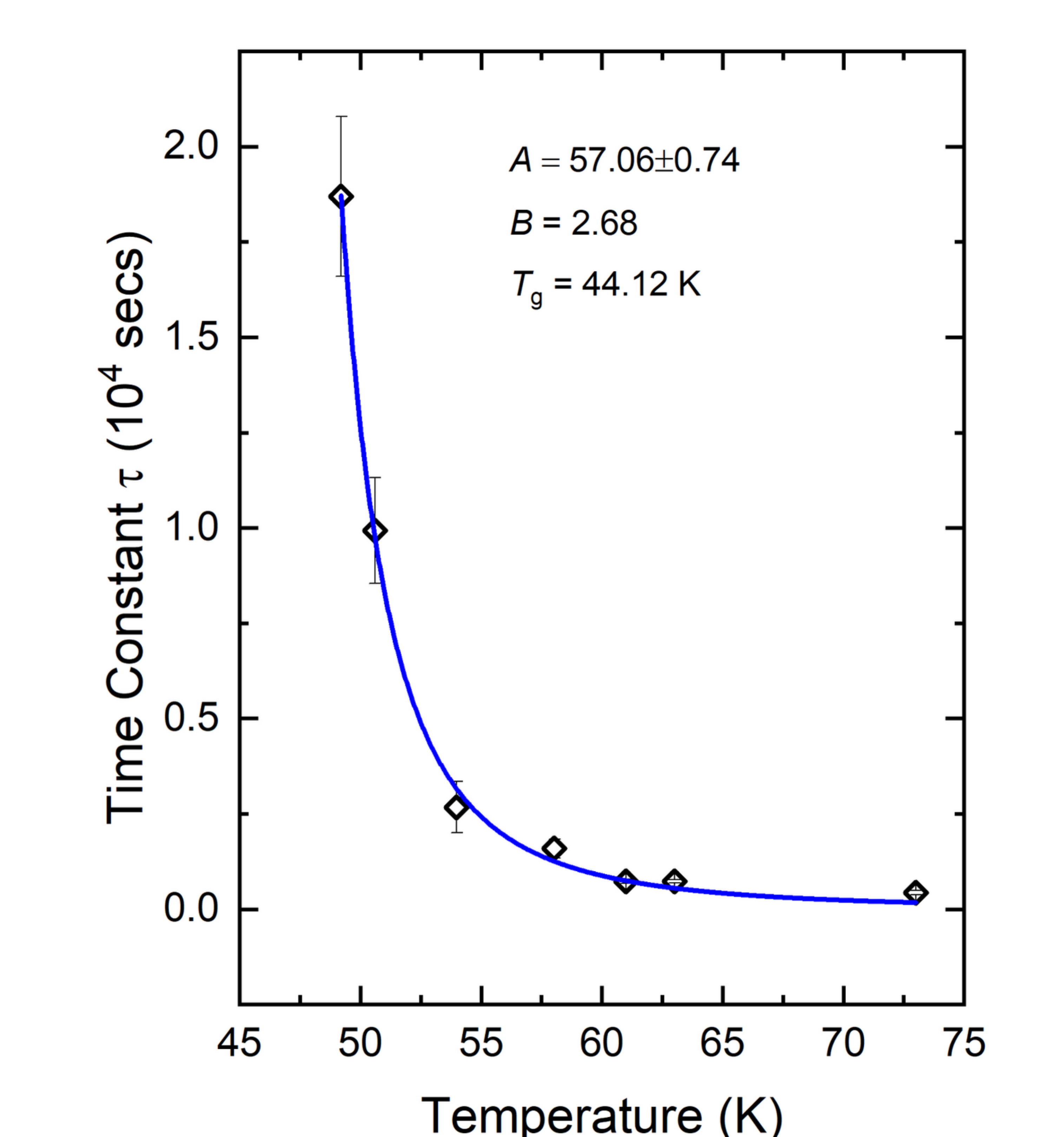}
 \end{center}
\caption{The power law fit to the deduced relaxation rate $\tau(T)$ with parameters given in the figure}
 \label{Fig:S6} 
\end{figure}

\newpage
 \noindent
{\bf f: Power law Fit for $\tau(T)$}\\
As mentioned in the main text, $\tau(T)$ can be fitted equally well to a power law of the form
\begin{equation*}
\tau_{0}(T)=\frac{A}{\left(T-T_{\mathrm{g}}\right)^{B}} \tag{15}
\end{equation*}
\noindent
This fit is shown is shown in Fig. (\ref{Fig:S6}). The fit is as good as to the Vogel-Fulcher law. The fit yields values of 44.12 K for $\mathrm{T}_{\mathrm{g}}$ and $\mathrm{B}=2.68$. As already mentioned the value of B does not agree numerical critical exponent (zv) observed in  computer simulations  and nonlinear susceptibility measurements to be $\sim 7$. It should be noted however that those simulations were for a nearest-neighbor random $\pm \mathrm{J}$ exchange Ising model on a cubic lattice. On the other hand, the mean field value of (zv) is 2 and values even lower have been quoted experimentally, in references given in the main text.

\end{document}